\documentclass[trackchanges]{aastex701}

\usepackage[nolist]{acronym}
\usepackage{comment}
\usepackage{}

\begin{document}
\correspondingauthor{Tremblay, C.D.}
\email{ctremblay@seti.org}

\title{A search for narrowband technosignatures from LTT 3780 with \\ the Allen Telescope Array and the Karl G. Jansky Very Large Array}

\author[0000-0002-4409-3515]{Tremblay, Chenoa D.}
\affiliation{These two authors contributed equally to this work}
\affiliation{SETI Institute, 339 Bernardo Ave, Suite 200, Mountain View, CA 94043, USA}
\affiliation{Berkeley SETI Research Center, University of California, Berkeley, CA 94720, USA}
\affiliation{Department of Physics and Astronomy, University of New Mexico, Albuquerque, NM 87131, USA}
\affiliation{National Radio Astronomy Observatory, 1003 Lopezville Rd., Socorro, NM 87801, USA} 
\email{ctremblay@seti.org}

\author[0000-0001-7057-4999]{Sheikh, Sofia Z.}
\affiliation{These two authors contributed equally to this work}
\affiliation{SETI Institute, 339 Bernardo Ave, Suite 200, Mountain View, CA 94043, USA}
\affiliation{Berkeley SETI Research Center, University of California, Berkeley, CA 94720, USA}
\email{fakeemail2@google.com}

\author[0000-0002-8604-106X]{Gajjar, Vishal}
\affiliation{SETI Institute, 339 Bernardo Ave, Suite 200, Mountain View, CA 94043, USA}
\email{fakeemail3@google.com}

\author[0000-0002-4869-000X]{Madhusudhan, Nikku }
\affiliation{Institute of Astronomy, University of Cambridge, Madingley Road, Cambridge CB3 0HA, UK}
\email{fakeemail3@google.com}

\author[0009-0003-9610-5068]{Gerrard, Isabel}
\affiliation{Breakthrough Listen, University of Oxford, Department of Physics, Denys Wilkinson Building, Keble Road, Oxford, OX1 3RH, UK}
\affiliation{SETI Institute, 339 Bernardo Ave, Suite 200, Mountain View, CA 94043, USA}
\email{izzy.gerrard@physics.ox.ac.uk}

\author[0000-0002-8071-6011]{Czech, Daniel}
\affiliation{Berkeley SETI Research Center, University of California, Berkeley, CA 94720, USA}
\email{fakeemail3@google.com}

\author[0000-0003-0804-9362]{Myburgh, Talon}
\affiliation{Mydon Solutions (Pty) Ltd., 102 Silver Oaks, 23 Silverlea Road, Wynberg, Cape Town, South Africa, 7800}
\affiliation{SETI Institute, 339 Bernardo Ave, Suite 200, Mountain View, CA 94043, USA}
\email{fakeemail3@google.com}

\author[0000-0001-6950-5072]{MacMahon, David E. }
\affiliation{Berkeley SETI Research Center, University of California, Berkeley, CA 94720, USA}
\email{fakeemail3@google.com}

\author[0009-0001-8677-372X]{Donnachie, Ross A. }
\affiliation{Mydon Solutions (Pty) Ltd., 102 Silver Oaks, 23 Silverlea Road, Wynberg, Cape Town, South Africa, 7800}
\affiliation{SETI Institute, 339 Bernardo Ave, Suite 200, Mountain View, CA 94043, USA}
\email{fakeemail3@google.com}

\author[0000-0003-2828-7720]{Siemion, Andrew P.V. }
\affiliation{SETI Institute, 339 Bernardo Ave, Suite 200, Mountain View, CA 94043, USA}
\affiliation{Breakthrough Listen, University of Oxford, Department of Physics, Denys Wilkinson Building, Keble Road, Oxford, OX1 3RH, UK}
\email{fakeemail3@google.com}

\author[0000-0002-7042-7566]{Lebofsky, Matthew}
\affiliation{Berkeley SETI Research Center, University of California, Berkeley, CA 94720, USA}
\email{fakeemail3@google.com}

\author[0000-0002-3430-7671]{Pollak, Alex W.}
\affiliation{SETI Institute, 339 Bernardo Ave, Suite 200, Mountain View, CA 94043, USA}
\email{apollak@seti.org}

\begin{abstract}
The LTT~3780 system hosts two known exoplanets—LTT~3780~b, a rocky super-Earth, and LTT~3780~c, a temperate sub-Neptune—orbiting a nearby M dwarf on opposite sides of the radius valley. LTT~3780~c has been proposed as a candidate Hycean world, making the system an important target for astrobiological investigation, particularly in light of recent JWST atmospheric observations. Although biosignature and technosignature searches both seek evidence of life beyond Earth, these approaches have historically been pursued independently. Well-characterized exoplanet systems provide an opportunity to combine these complementary search strategies. In this work, we conducted radio technosignature observations of the LTT~3780 system using both the Allen Telescope Array (ATA) and the Karl G. Jansky Very Large Array (VLA). The two facilities provide complementary observational capabilities, with the ATA optimized for wide-band multi-beam post-processing analyses and the VLA enabling high-sensitivity real-time interferometric searches. Across $\sim$30 hr of total observing time, we searched for narrowband Doppler-drifting signals in the frequency range $\sim$1–10 GHz. After applying comprehensive radio-frequency interference mitigation and multi-beam consistency tests, no candidate signals consistent with astrophysical or technosignature origins were identified. We place minimum detectable effective isotropic radiated power limits of $4.7\times10^{12}$–-$3.6\times10^{13}$\,W across the observed bands and facilities. Although no technosignatures were detected, this work demonstrates how complementary observation and analysis strategies can be applied to exoplanets of astrobiological interest and serves as a pathfinder for future combined biosignature and technosignature investigations.

\end{abstract}


\section{Introduction} 
\label{sec:intro}

The search for life beyond Earth has traditionally proceeded along two largely independent pathways: the search for metabolic biosignatures that trace biological activity and the search for technosignatures that may indicate the presence of technologically capable civilizations (also known as the search for extraterrestrial intelligence or SETI). Recent advances in exoplanet discovery and characterization, particularly for rocky planets orbiting nearby stars, motivate a more unified approach: we can now incorporate context-specific information \citep[e.g., ][]{meadows2018exoplanet, chan2019deciphering} to refine both biosignature and technosignature searches on the same target. Exoplanet spectra can be used to identify potential biosignatures, such as atmospheric chemical disequilibria \citep{krissansen2018disequilibrium}, while also accounting for surface–atmosphere interactions and global redox processes that link the biosphere, atmosphere, and hydrosphere. Meanwhile, technosignature searches probe a complementary regime in which technological activity, coupled to these other systems through the \textit{technosphere} \citep{zalasiewicz2017scale}, could modify the electromagnetic or chemical environment of a planet. Potential technosignatures may also include industrial atmospheric pollutants or other anthropogenic chemical disequilibria detectable in exoplanet spectra \citep{lin2014detecting, schneider2010far}, as well as the presence of technology. The technosphere itself, if detected, can even serve as a tracer of other planetary properties \citep[e.g., planetary rotation][]{sullivan1978eavesdropping}. When applied jointly, biosignature and technosignature approaches offer a more complete framework for assessing habitability and the presence of life, especially for terrestrial planets, where geological, biological, and technological processes may be tightly coupled.

We therefore take advantage of this emerging era in which the astrobiology and SETI communities can work in tandem to investigate exoplanets, the stellar systems that contain them, and their potential habitability. As an initial step in this collaborative effort, we previously observed K2--18b -- a proposed Hycean world \citep{Madhusudhan_2021} -- using the MeerKAT telescope in South Africa and the Karl G. Jansky Very Large Array (VLA) in New Mexico \citep{tremblay_k218b}. Although no candidate signals remained after applying a comprehensive mitigation pipeline to remove \ac{RFI}, that study established a framework for conducting coordinated searches for technosignatures in systems of astrobiological interest.

Recent JWST observations of the LTT~3780 system further motivate its study in an astrobiological context. In particular, the planet LTT~3780~c has been predicted to be a possible Hycean world \citep{Madhusudhan_2021}. Using JWST transmission spectroscopy, \citet{Rigby_2025} report a moderately-strong detection of CH$_4$ along with potential trace detections of hydrocarbons and sulfur-bearing species, including isobutylene, propylene, and/or dimethyl sulfide, at abundances of $\sim$1–-100\,ppmv in the atmosphere of LTT~3780~c. Although these molecules may not be uniquely indicative of biological or technological processes, their possible presence highlights a chemically active environment that warrants further investigation. In systems such as LTT~3780, where both a rocky super-Earth (LTT~3780~b) and a sub-Neptune companion are present (LTT~3780~c), combined studies of atmospheric composition and radio technosignatures provide complementary constraints on planetary processes. Technosignature searches offer an independent pathway to probe the presence of technological activity that may not be reflected in atmospheric chemistry alone.

Although the potential habitability of Hycean and volatile-rich sub-Neptune worlds remains uncertain, such environments broaden the range of planetary conditions under consideration for astrobiology and technosignature studies. Proposed scenarios include habitable ocean layers beneath hydrogen-rich atmospheres \citep{Madhusudhan_2021}, as well as the possibility of technological activity associated with oceanic or non-surface biospheres \citep{Lingam_2023,Balbi_2024,Catling_2025}. Technosignature searches therefore provide a complementary probe that does not require strictly Earth-like planetary conditions or evolutionary pathways\footnote{For example, humans have generated significant artifact and radio technosignatures on Mars, a decidedly uninhabitable environment for biology alone.}. Moreover, with a broad search on the entire extra-solar system, we can be agnostic to the exact position a transmitter, which is often true of searches completed using radio telescopes (e.g \citealt{enriquez2017turbo,gajjar2019breakthrough}). 




In this work, we extend the framework previously applied to K2-18b \citep{tremblay_k218b} to the LTT~3780 system using complementary observations from the \ac{ATA} and VLA. The observations were not simultaneous; rather, the two facilities provide distinct and complementary search capabilities across overlapping regions of parameter space. The wide synthesized beams of both arrays encompass the full stellar system, enabling searches for technosignatures associated with both known planets as well as the surrounding environment. 

The ATA observations emphasize multi-beam candidate discrimination and post-processing analyses across broad fractional bandwidths, while the VLA observations leverage high-sensitivity interferometric beamforming and real-time Doppler-drift searches. For the ATA analysis, we additionally focus on signals consistent with expectations for LTT~3780~c, the $2.66\,M_{\oplus}$ exoplanet whose radius and density may be consistent with volatile-rich or Hycean-world interpretations \citep{Cloutier_2020,Nowak_2020,Rigby_2025}. 

Together, these complementary observing strategies allow us to probe a broader range of potential narrowband signal phenomenology than either facility alone. In this paper, we present results from ATA and VLA searches spanning multiple orbital phases, including planetary occultations, and covering a wide radio bandwidth. In Section~\ref{sec:observations}, we describe the observing strategies for both telescopes. In Section~\ref{sec:results}, we present the search methodologies and results from each facility. We discuss the implications of these results in Sections~\ref{sec:discussion} and \ref{sec:conclusion}.

\section{Observations}
\label{sec:observations}

\subsection{Determinations of Drift Rate and Orbital Event Timings}
\label{ssec:drift_rate_and_transits}

\subsubsection{Drift Rate}
\label{sssec:drift rate}
For this work, we focus primarily on the orbital parameters of LTT~3780~c, as this was the planet targeted by the JWST observations \citep{Rigby_2025}. However, we note that with the size of the point spread function of both telescopes, the entire planetary system is within the coherent beam. LTT~3780~c is expected to be tidally locked, with an orbital inclination of approximately $90^{\circ}$ and an orbital period of 12.25~days \citep[e.g.,][]{Cloutier_2020,Bonfanti_2024}. If a narrowband transmitter were associated with the planet or its local environment (i.e. a geocentric orbiting satellite), the received signal frequency would vary over time due to Doppler acceleration from the planet’s orbital motion \citep[e.g.,][]{sheikh2019, Li_2022}. Additional contributions to the observed drift rate arise from Earth’s rotation and orbital motion, while the contribution from the planetary rotation of LTT~3780~c is expected to be minimal because of tidal locking.

Using the system parameters for LTT~3780 ~ c from the NASA Exoplanet Archive\footnote{\url{https://exoplanetarchive.ipac.caltech.edu/}}, we estimate a maximum orbital drift rate contribution of approximately 1.38~nHz at 1 \, GHz \citep{sheikh2019,Li_2023}, which scales linearly with the observation frequency. By comparison, Earth’s rotation and orbital motion contribute drift rates of order $\sim$0.1 and $\sim$0.02~Hz\,s$^{-1}$, respectively, at 1\,GHz \citep[e.g.,][]{enriquez17}. To ensure sensitivity to both planet-associated and anthropogenic-style narrowband emitters, the ATA and VLA searches sampled substantially broader drift-rate ranges than those expected from the orbital dynamics of the system alone. The full drift-rate ranges searched for each facility are described in Sections 3.1 and 3.2.



\subsubsection{Orbital Events}
\label{sssec:orbital_events}
\begin{figure}
    \centering
    \includegraphics[width=1\linewidth]{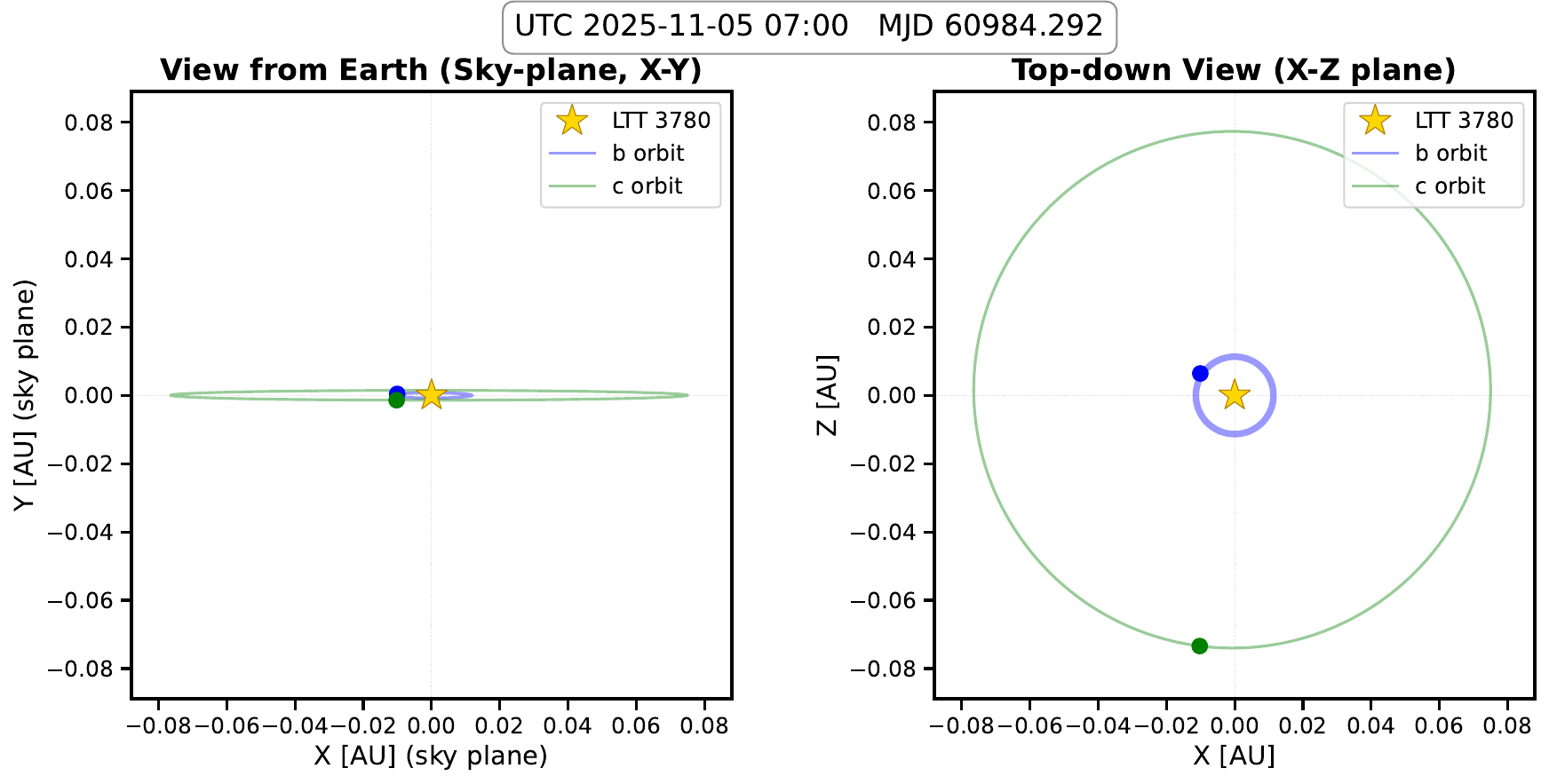}
    \caption{Geometry of the LTT 3780 planetary system at the time of our radio technosignature observations. The left panel shows the sky-plane (X–Y) view as seen from Earth, while the right panel shows the top-down (X–Z) view of the system. The orbits of planets b (blue) and c (green) are shown relative to the host star (yellow). Our observations from the ATA, mentioned in Table \ref{tab:ata_observing_log} on 5th November 2025, were timed to coincide with a planet–planet occultation (PPO), when planet b passes behind planet c along the line of sight, providing a unique configuration to search for radio technosignatures during mutual planetary alignment.}
    \label{fig:PPO_event}
\end{figure}
LTT~3780 is a system that hosts two planets with well-characterized orbits \citep{Cloutier_2020,Bonfanti_2024}. We can thus use the known orbital geometries in order to provide additional geometric and temporal constraints that may aid in the interpretation and localization of candidate signals. \citet{Gajjar_2026} showed that, near GHz frequencies, plasma-density fluctuations along the line of sight can broaden otherwise narrowband emission, with the magnitude of the effect depending on the stellar-wind environment, observing frequency, and star--planet--observer geometry. Thus, for a system such as LTT~3780, phase-resolved narrowband search framework provides unique opportunity to investigate such intrinsic effects.  
In addition, previous work has noted that technosignature surveys may be more likely to pick up signals from radio ``spillover'' from intra-system communication if those surveys are timed to coincide with \ac{PPO}s \citep[which can enhance detection probability by $4 \times 10^5$ in the analogous Earth-Mars case, ][]{fan2025detecting}. A few recent technosignature searches have been conducted during primary transit \citep[e.g., ][]{sheikh2023green}, secondary transit \citep{Barrett_2025}, or aligned with \ac{PPO}s \citep[][]{Tusay_2024} but timing survey observations to coincide with orbital events in known exoplanetary systems is still a relatively novel technique. We note that although the current habitability of the known planets in LTT~3780 remains uncertain, PPO observations are still motivated by geometric considerations rather than by the assumption that either planet hosts a surface civilization. If technological activity exists anywhere in the system—on a planet, in orbit, on moons, probes, stations, or other infrastructure—planet–planet alignments may increase the probability that our line of sight intersects radio leakage from intra-system communication or navigation links.

Given that these two planets are neither resonant nor especially tightly packed, we can propagate their Keplerian orbits forward in time to predict the timings of transits and \acp{PPO}. We adopted system ephemerides from the NASA Exoplanet Archive, using the updated orbital parameters reported by \citet{Bonfanti_2024}. We then propagated the orbits of planets~b and~c with a Keplerian model anchored to their published mid-transit epochs. At each time step, we mapped the instantaneous planetary positions into the sky-plane $(X,Y)$ and searched for intervals in which the projected $(X,Y)$-coordinates of the two planets agreed within one Jupiter radius ($\approx 4.8\times10^{-4}$~AU). Contiguous intervals satisfying this criterion were classified as candidate \ac{PPO} windows and used to guide the scheduling of our ATA observations. {\bf Figure \ref{fig:PPO_event} shows one such configuration.} From \citet{Bonfanti_2024}, the primary and secondary transit durations for LTT~3780~b and LTT~3780~c range from $\approx 47.9 \pm 0.7$ minutes to $92.5 \pm 1.6$ minutes, while the candidate PPO intervals typically span 10--30 minutes (see Table \ref{tab:ata_observing_log}), depending on the projected orbital geometry. It should be noted that the timing uncertainties, of order $\sim$1 minute, are small compared to the observing windows, as shown in Table~\ref{tab:ata_observing_log} and therefore do not significantly affect the interpretation of the orbital configurations or the associated signal-localization tests.

\subsection{Allen Telescope Array}
\label{ssec:ata_obs}

The \ac{ATA}, located at \ac{HCRO} in Northern California, is a 42-element radio interferometer consisting of 6.1-m dishes, with 28 of 42 currently refurbished with cryocooled dual-polarization log-pyramidal ``Antonio'' feeds \citep{welch2017new}. It is capable of recording simultaneously from four independent 672\,MHz-wide tunings. The \ac{ATA} conducted four observations of LTT~3780 over roughly 20~hours, split between four observing sessions from November to December 2025, covering 2.7\,GHz of bandwidth. Observations of LTT~3780 used the active 28-dish array and consecutive 672 MHz-wide tunings to record a continuous bandwidth spanning 1000--3688~MHz (i.e. center frequencies of 1336, 2008, 2680, and 3352\,MHz). The bandwidth observed with the ATA covered all of L-band (1--2\,GHz) and most of S-band (2-4 GHz).

Each observational session was preceded by a 10~minute calibration scan of 3C48 \citep[a bright flux calibrator source with a well-characterized flux density from][]{perley2017accurate} observed using the \ac{ATA} continuum correlator mode. This calibration, processed utilizes an in-house CASA pipeline \citep{casa2022casa} to ensure correct phasing of the beamformer and provided estimates of system sensitivity.

Data were then recorded using the \ac{BLADE} beamformer\footnote{\url{https://github.com/luigifcruz/blade}} backend to synthesize two beams on the sky: the ``on-beam'' was centered on LTT~3780 (RA:10:18:34.78, Dec:--11:43:04.08), which itself was placed at boresight of the incoherent beam, and the ``off-beam'' was placed five synthesized beamwidths away. This on-off coherent beamforming strategy can be used by radio arrays performing technosignature searches to leverage interferometry for sky localization while keeping data rates manageable \citep[e.g., ][]{Tusay_2024}, and is analogous to the ``nodding'' or ABACAD on-off strategy for sky localization used by single dish telescopes \citep[e.g., ][]{painter2025novel}. By delaying the signals from different dishes in the array, the \ac{ATA} is able to synthesize two or more simultaneous ``coherent'' beams whose beam-widths are significantly narrower on the sky than the primary beam from a single array element. Discriminating signals present only in the coherent on-beam from those in both coherent beams serves as a filter for signals originating from the direction of the target-of-interest, in this case LTT~3780 (see Section~\ref{ssec:ata_results}). The frequency resolution and sampling times (3.81\,Hz and 8.4\,s respectively) were set to sufficiently resolve narrowband signals at the expected drift rate of 5.1\,Hz\,s$^{-1}$ at the highest frequencies (see Section~\ref{ssec:drift_rate_and_transits}), and the inter-beam distance was calculated for the center frequency of each independent tuning. 

The four observing sessions, as shown in Table \ref{tab:ata_observing_log}, included seven notable orbital configurations of the LTT~3780 system: one primary transit of LTT~3780~c, two secondary transits of LTT~3780~c, and four \ac{PPO}s between LTT~3780~b and LTT~3780~c (see Table~\ref{tab:ata_observing_log}). Each transit lasted approximately 2~hours. Only the first $\sim$78\% of the observed primary transit was captured due to ``windsocking" (i.e., the array automatically halted the recording during the transit due to dangerous high-wind conditions). Additionally, the latter $\sim$4.5 hours of data from the 2025-12-18 observation were missing data between 2680-2776 MHz due to a compute node failure in data-recording of a 96\,MHz sub-band; the other 27 nodes were processed in the same fashion as the rest of the recordings.  

\begin{deluxetable*}{cccccc}
\tablecaption{Observing sessions on the \ac{ATA} for this campaign.\label{tab:ata_observing_log}}
\tablewidth{0pt}
\tablehead{
\colhead{Date} &
\colhead{Event(s)} &
\colhead{Obs.\ Start (UT)} &
\colhead{Obs.\ End (UT)} &
\colhead{PPO Start (UT)} &
\colhead{PPO End (UT)}
}
\startdata
2025-11-05 & Primary Transit \& PPO & 11:48 & 14:18 & 13:09 & 13:31 \\
2025-11-11 & Secondary Transit \& PPO & 11:25 & 17:20 & 16:03 & 16:13 \\
2025-12-12 & PPO & 09:23 & 15:13 & 12:44 & 14:54 \\
2025-12-18 & Secondary Transit \& PPO & 09:02 & 14:57 & 12:20 & 12:30 \\
\enddata
\tablecomments{Observing intervals are anchored to the scheduled start times in the campaign log. PPO windows are the predicted sky-plane alignment intervals ($|X_{\rm b}-X_{\rm c}|<1\,R_{\rm Jup}$) from our Keplerian model for the corresponding calendar date. These observations comprise $\sim$20.5 hours and seven LTT~3780 orbital configurations of note. Orbital events are reported relative to LTT~3780~c.}
\end{deluxetable*}

\subsection{Very Large Array}
\label{ssec:vla_obs}

The LTT~3780 system was observed with the \ac{VLA} across three epochs and three frequency bands (Table~\ref{tab:obs_setup_cosmic}), with the phase center set to RA = 10$^{\mathrm h}$18$^{\mathrm m}$34.78$^{\mathrm s}$, Dec = $-11^\circ$43$'$4.08$''$, the position of the M-dwarf star LTT~3780. Data were recorded and processed using the Commensal Open-Source Multimode Interferometer Cluster (COSMIC; \citealt{Tremblay_cosmic}), an Ethernet-based digital backend that captures a copy of the antenna voltage streams, applies calibration, forms coherent beams toward selected targets, and searches for narrowband signals exhibiting Doppler drift. Each detected signal (hit) was logged into an SQL database and a calibrated voltage segment (“postage stamp”) surrounding the detection was retained for post-processing and evaluation. The observations used 8-bit analogue-to-digital sampling. At these bit depths, the sensitivity loss introduced by signal quantization is negligible for the technosignature searches considered here and therefore has little impact on the sensitivity determined by the radiometer equation \citep[e.g.,][]{tremblay_k218b}. Calibration data and derived solutions were also preserved for subsequent analysis. Additional details of the COSMIC system and pipeline are provided in \citet{Tremblay_cosmic,Tremblay_VLASS}.

Although COSMIC typically operates in a commensal mode, the observations analyzed here were obtained through approved observing time (Proposal ID: VLA/25B-235). The observations were conducted at S-band (2–4 GHz), C-band (4–8 GHz), and X-band (8–12 GHz), with COSMIC recording up to 1.4\,GHz of instantaneous bandwidth in each session. Each observing block began with a 5-minute scan of the calibrator 3C286 to establish bandpass calibration and system configuration, using standard radio interferometry techniques. Data were recorded simultaneously by both the VLA WIDAR correlator and COSMIC; however, only the COSMIC data products were used in this analysis.

The calibration quality metrics ranged from 0.90 to 0.98 across 21 operational antennas, consistent with stable phase calibration and reliable coherent beamforming performance.

\begin{deluxetable*}{lccccc}
\tablecaption{Parameters for the COSMIC observations and real-time pipeline.}
\tablewidth{0pt}
\tablehead{
\colhead{Band} &
\colhead{Date} &
\colhead{Start Time} &
\colhead{Planet b} &
\colhead{Planet c} &
\colhead{Polarizations} \\
&
\colhead{(UTC)} &
&
\colhead{Phase} &
\colhead{Phase} &
\\
}
\startdata
S  & 2025-11-23 & 16:03  & 0.576392 & 0.478803 & 2 \\
S  & 2025-12-02 & 14:58  & 0.230612 & 0.209676 & 2 \\
S  & 2025-12-10 & 13:35  & 0.567123 & 0.857911 & 2 \\
C  & 2025-11-19 & 13:45  & 0.245909 & 0.144511 & 2 \\
C  & 2025-11-21 & 16:02  & 0.972607 & 0.315511 & 2 \\
C  & 2025-11-23 & 15:40  & 0.555605 & 0.477499 & 2 \\
X  & 2025-11-23 & 15:17  & 0.534818 & 0.476196 & 2 \\
X  & 2025-12-01 & 15:20  & 0.949054 & 0.129305 & 2 \\
X  & 2025-12-03 & 10:39  & 0.297974 & 0.276613 & 2 \\
\enddata
\tablenotetext{}{All observations used 21 antennas, a 10$\sigma$ threshold, integration time of 56\,s, a channel width of 2\,Hz and 8-bit recording.}
\label{tab:obs_setup_cosmic}
\end{deluxetable*}


\subsection{Comparison of Observations}
The ATA and VLA observations employed different signal-processing architectures optimized for their respective systems. At the ATA, candidate detections were processed through the ATSAT pipeline, which performs post-processing candidate aggregation and RFI rejection using simultaneous “on” and “off” coherent beams. Additionally, the ATA records and processes the entire bandwidth simultaneously across the four tunings. In contrast, the VLA/COSMIC observations used \textsc{seticore}, a GPU-accelerated real-time Doppler-drift search engine operating directly on beamformed voltage data. Each frequency band was observed at different times and dates, as listed in Table \ref{ssec:vla_obs}.

Although the VLA analysis did not explicitly define dedicated “off beams,” up to four simultaneous coherent beams were formed across the field of view, enabling multi-beam consistency tests analogous to the off-beam rejection methods employed at the ATA. Signals detected simultaneously across multiple beams were treated as likely interference or instrumental artifacts rather than sky-localized technosignatures.

\section{Results}
\label{sec:results}

\subsection{Allen Telescope Array}
\label{ssec:ata_results}

We processed all of the data with \ac{ATSAT}---an efficient narrowband, Doppler-drifting signal search pipeline on the \ac{ATA}. \ac{ATSAT} integrates the software packages \texttt{bliss}\footnote{\url{https://github.com/UCBerkeleySETI/bliss} \citep{nathan_west_2025_bliss}.} and \texttt{NBeamAnalysis}\footnote{\url{https://github.com/isabelgerrard/NBeamAnalysis}  \citep{ntusay_igerrard_nbeam}.} with custom \ac{RFI} ``blanking'' and plotting functionality (See the Appendix and Table \ref{tab:blanking_ranges} for the exact blanking channels). \texttt{bliss} detects Doppler-drifting narrowband signals (``hits''), and \texttt{NBeamAnalysis} spatially compares hits between beams. \texttt{bliss} was applied to search for narrowband hits with drift rates $\pm$5.3\,Hz\,sec$^{-1}$ and a \ac{SNR} threshold of 15. \textit{NBeamAnalysis} was run with a `spatial filter' threshold of 5.29, determined using $\sqrt{N_{antennas}}$ = $\sqrt{28}$ = 5.29 as a rough estimate of the minimum expected \ac{SNR} difference between synthesized beams for a truly sky-localized signal. Any hits that are nearly identical in frequency and have an SNR-ratio less than the spatial filter threshold ($<$5.29) between beams are excluded from the next steps. For the remaining hits, \texttt{NBeamAnalysis} calculates the \ac{SNR}-ratio and dot product (``DOT score'') between beams and plots hits with an \ac{SNR}-ratio greater than 5.29 for visual inspection. A high \ac{SNR}-ratio indicates that the signal was brighter in the on-beam, more consistent with a point source from the target location on the sky. 

To remove \ac{RFI} that is local to the telescope, we identified ``blanking ranges'' by visually identifying intervals at least 0.5\,MHz wide with a high density of hits across all drift rates searched ($\pm$5.3\,Hz\,s$^{-1}$) which indicates \texttt{bliss} could not match the underlying signal to a single drift rate (suggesting local interference). All hits in the ``frequency-blanking ranges'' listed in Table \ref{tab:blanking_ranges} were removed from further processing and outputs (i.e.``blanked") in order to filter out known or suspected \ac{RFI} regions, which removed 27\% of the band. This is consistent with previous surveys at \ac{HCRO}, as it is not in a radio quiet zone and L-band and S-band contain the highest amounts of Earth-based transmitters in the cm-wave spectrum.

Processing of the entire $\sim$20 hours of data ($\sim$58~TB) was completed by \ac{ATSAT} in $\sim$6.6 days and resulted in 6,842 plots for visual inspection. The frequency-blanked regions (27\% of the observed bandwidth) were responsible for an average of 82\% of hits detected by \texttt{bliss} per night, confirming that these regions were indeed \ac{RFI}-heavy.

\begin{deluxetable}{ccccccc}
\tablecaption{Data from the ATA processed using the ATSAT pipeline.\label{tab:atsat_processing}}
\tablehead{
\colhead{Observation} &
\colhead{Processing Time} &
\colhead{\textit{bliss} hits} &
\colhead{Hits blanked} & 
\colhead{Hits after blanking} &
\colhead{Hits after spatial filtering} &
\colhead{Output plots} \\
&
\colhead{(h)} &
\colhead{(Count)} &
&
\colhead{(Count)} &
\colhead{(Count)} &
\colhead{(Count)}
}
\startdata
2025-11-05 & 15.4 & 30945821 & 84.88\% & 4676531 & 1843435 & 493 \\
2025-11-11 & 45 & 81396959 &  83.61\% & 13340602 & 5214573 & 937 \\
2025-12-12 & 45 & 86037822 & 82.30\% & 15227609 & 5853055 & 2414 \\
2025-12-18 & 53.9 & 83640177 & 79.05\% & 17522429 & 7142153 & 2998 \\
\hline
Total & 159.3 & 282020779 & 82\% & 50767171 & 20053216 & 6842 \\
\enddata
\label{tab:atsat_processing}
\end{deluxetable}

\begin{figure}
    \centering
    \includegraphics[width=\linewidth]{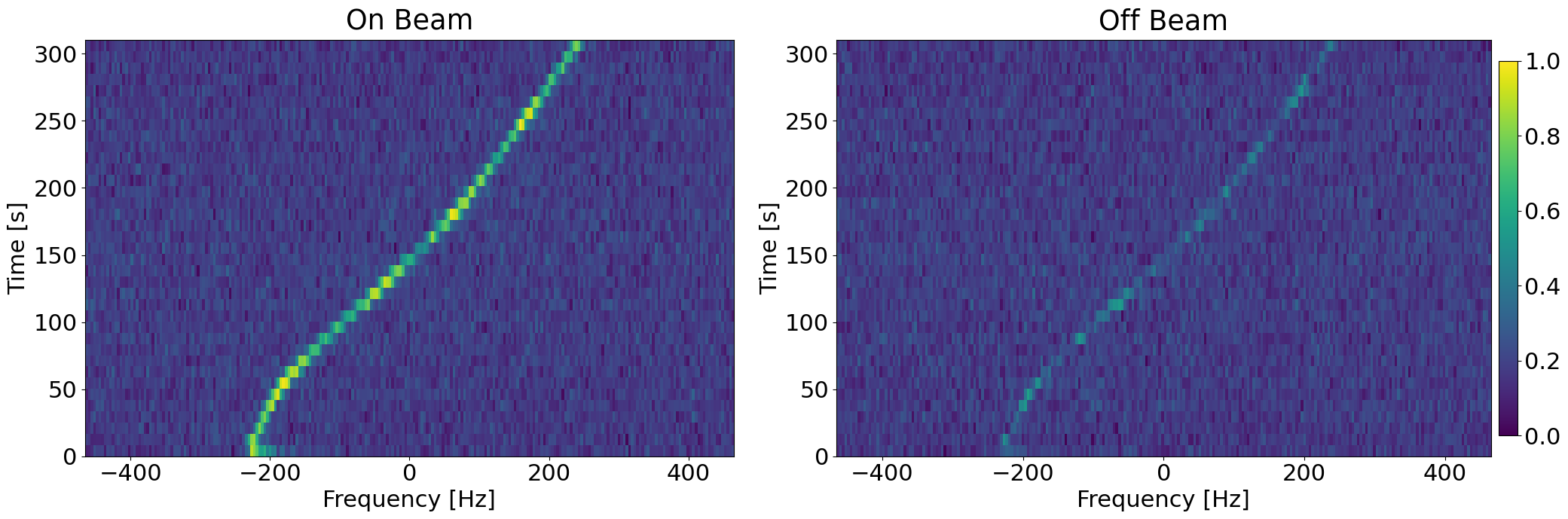}
    \caption{An output plot from the ATSAT pipeline depicting two plots of a signal which was flagged for further investigation after visual inspection. The plotted signal corresponds to hit with a drift rate of 1.326 Hz/s and an \ac{SNR} ratio of 5.45. The plots show frequency offset from a center frequency of 2400.014593 MHz, time on the y-axis, and normalized intensity (scaled to the brighest and faintest intensities in the leftmost panel) in the colorbar. The left panel is the dynamic spectrum of the ``on-beam'' which was pointed towards LTT 3780, while the right panel shows the dynamic spectrum of the ``off-beam'' which serves as the control. This event is definitively technological due to its narrowband nature, is either inherently changing frequency or Doppler drifting, and is fainter in the off-beam, indicating potential sky-localization. This signal was determined to be \ac{RFI} due to its varying \ac{SNR} ratio displayed in Figure~\ref{fig:ATA_event_stacked}.}
    \label{fig:ATA_event}
\end{figure}

\begin{figure}
    \centering
    \includegraphics[width=0.8\linewidth]{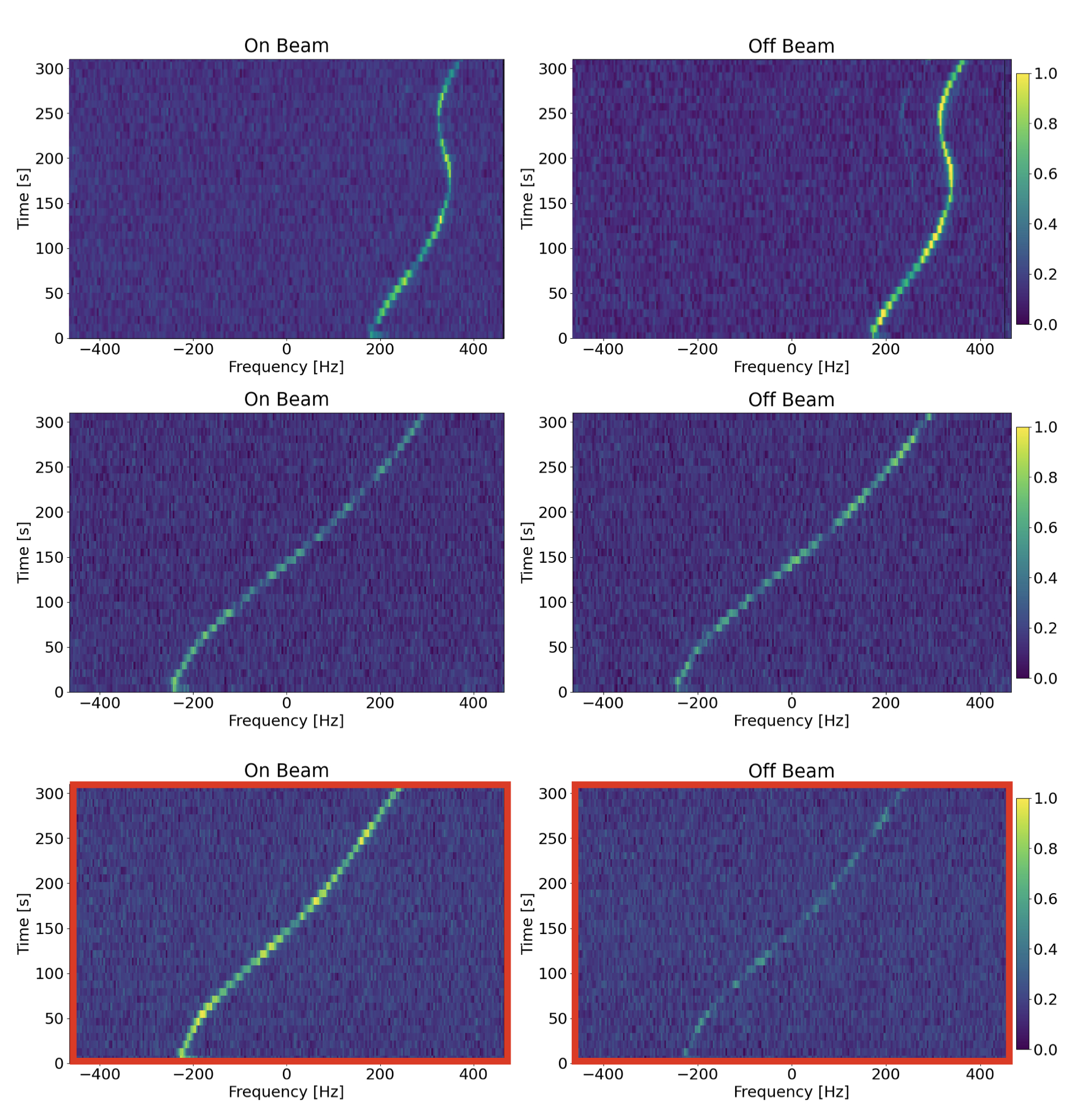}
    \caption{The event from Figure~\ref{fig:ATA_event} centered to 2400.014593 MHz (bottom row) shown below two prior five minute scans of the same target in the same frequency range (top two rows). All three rows are dynamic spectrum pairs with the same characteristics as Figure~\ref{fig:ATA_event}, with all six subplots normalized to the on-beam of the bottom row. The first row was taken 18min 34s before the bottom row, and the middle row was taken 6min 11s before the bottom row. Note that the signal is present in all scans with varying start frequency and morphology, but that the intensity difference between the on- and off-beams is different between the bottom row and the top two rows: slightly brighter in the off-beam rather than the on-beam. A wobbling narrowband signal of this nature could be due to a periodic local \ac{RFI} source or an artifact internal to the \ac{ATA} system; regardless of source, it is definitively not originating from the sky position of LTT~3780.}
    \label{fig:ATA_event_stacked}
\end{figure}

We visually inspect each of the 6842 graphs from the total output plots column of Table~\ref{tab:atsat_processing}. Three of these 6842 events (one of which is shown in Figure~\ref{fig:ATA_event}) were flagged for further investigation due to their slightly above-threshold \ac{SNR}-ratios (indicating potential consistency with a point source) and their narrow frequency widths (indicating technological origin). In this campaign, we took repeated 5 minute scans of the same target for several hours in each frequency band. This allows us to investigate the frequency range where the three events appear before and after the scan where the \texttt{bliss} hit was triggered. For a true, sky-localized point source we would expect the \ac{SNR} ratio to stay relatively consistent from scan to scan---perhaps slowly changing over time with the beam pattern, but \textit{always} being brighter in the on-beam than in the off-beam. However, in each of the three events, the scans before and after the scan with the target event show widely-varying SNR ratios where the signal flips from being brighter in the on-source to being brighter in the off-source (an example is shown in Figure~\ref{fig:ATA_event_stacked}). This implies that all three events are not consistent with a point source at the location of LTT~3780 and could even be attributed to \ac{RFI} coming into the sidelobes.

The example in Figure~\ref{fig:ATA_event_stacked} is particularly eye-catching due to having a very similar morphology in two neighbouring scans. We found at least 22 instances of hits with nearly the same start frequency and a similar range of detected drift rates. We then visualized an additional five scans surrounding the event (two of which are shown for example in the top and middle rows of Figure~\ref{fig:ATA_event_stacked}) and found that the signal varies in start frequency within a few hundred Hz and has different drift characteristics over time, but always starts at zero drift in each scan and drifts to higher frequencies. This behaviour has two common potential explanations: 1) the source of the signal is an artifact or instrumental effect internal to the \ac{ATA}'s digital signal processing, causing the signal to reset to a similar location at the start of the scan 2) the signal is external to the system but periodically wandering with a period close to 360 seconds, aligning near the scan cadence \citep[this is seen occasionally in Breakthrough Listen Green Bank Telescope data from Iridium satellite transmissions e.g., Figure 4d from][]{sheikh2023green}).

Regardless of source, the original event has a beam-ratio of 5.45 (brighter in the on-beam) but many of the similar hits have beam-ratios of 0.64 and 0.94 respectively for the top and middle rows; (brighter in the off-beam in both cases), implying that the signal cannot be sky-localized and must be some kind of local interference (whether internal or external to the telescope). This kind of interferer has not been reported before in published \ac{ATA} datasets, but is minimally problematic for this survey (and would be similarly minimally problematic to future work) due to its limited presence in the dataset (only one instance needing additional context out of over 6000 events in 20 hours).

Given that no technosignatures were found in the \ac{ATA} data, upper limits of the \ac{EIRP} were determined using the same methods as \citet{Tusay_2024}, and are listed in Table \ref{tab:eirp}.

\subsection{Very Large Array}
\label{ssec:vla_results}
To remove signals likely associated with \ac{RFI}, we follow the post-processing pipeline described in \cite{tremblay_k218b} and the known RFI is outlined in the Appendix (Table \ref{tab:blanking_ranges_cosmic}). This process filters out signals associated with known \ac{RFI}, and subsequently applies additional filters to select: signals within expected drift rate ranges; signals with \ac{SNR}s between 10 and 1000; signals that appear only toward the LTT~3780 system; and signals that do not recur at consistent frequency and drift rate on other observing days. All remaining unique signals are then visually inspected using visualization tools within the \textsc{seticore} pipeline \footnote{\url{https://github.com/lacker/seticore}} to verify consistency with astrophysical or technosignature-like origins to identify candidates that warrant further analysis.
\subsubsection{S-band Data}
\label{sssec:vla_s}

As described in \citet{Tremblay_VLASS}, the S-band data was passed through a filter within the real-time pipeline for \ac{RFI} pre-processing. Therefore, the first filter to remove the \ac{RFI} only removed an additional 0.01\% of the signals (see Table \ref{tab:cosmic_processing}). By filtering the remaining signals using a drift rate cut of $\pm$4.1\,Hz\,s$^{-1}$ and a \ac{SNR} window of 10 to 1000, only 21\% of the originally detected signals remained. Of the 21,151 signals, only 5084 were constrained to the coherent beam pointing toward LTT~3780 and of those 197 signals were unique toward the position (not present in the other coherent beams). Of the 197 signals, 19 matched signals with the same frequency and drift rates contained within the incoherent beam and coherent beams, but all showed a ratio of unity, suggesting they are local \ac{RFI}.

Thus, a total of 178 signals were visually inspected to evaluate whether they were time-persistent, exhibited consistent frequency drift, and were present in all operational antennas. Examination of the beamformed dynamic spectra identified only three signals that appeared time-persistent; however, none of these were detected in all antennas. We therefore conclude that all detected signals are attributable to interference and do not represent technosignatures associated with LTT~3780 or the associated exoplanets. The upper limits on the \ac{EIRP} as determined using the equations in the Appendix of \cite{tremblay_k218b}, are listed in Table \ref{tab:eirp}.

\begin{deluxetable}{lccccc}
\tablecaption{Data from the COSMIC processed using the realtime pipeline.\label{tab:cosmic_processing}}
\tablehead{
\colhead{Band}&
\colhead{Observation} &
\colhead{\textit{seticore} hits} &
\colhead{\% Hits blanked} & 
\colhead{Hits after blanking} &
\colhead{Hits after spatial filtering} \\
&
 &
\colhead{(Count)} &
&
\colhead{(Count)} &
\colhead{(Count)} 
}
\startdata
S&2025-11-23 & 30781  & 0.01 &  30769& 78   \\
S&2025-12-02 & 47438  &  0.00 & 47438&  65  \\
S &2025-12-10 &  42895 & 0.01&  42895& 54   \\
C& 2025-11-19 & 109694 &  80.25&  21662& 47   \\
C& 2025-11-21 & 157153 &  84.90&  23727& 73   \\
C& 2025-11-23 & 109783 &  80.11&  21833& 57   \\
X& 2025-11-23 & 18359 &  25.37& 13702 & 74   \\
X& 2025-12-01 & 81250 &  10.58&  72657& 133   \\
X& 2025-12-03 & 26816 &  47.83&  13989& 19   \\
\enddata
\label{tab:atsat_processing}
\end{deluxetable}


\subsubsection{C-band Data}
\label{sssec:vla_c}

As noted by the \ac{NRAO} \ac{RFI} observers information\footnote{\url{https://science.nrao.edu/facilities/vla/observing/RFI/C-Band}}, C-band possesses significant sources of \ac{RFI}. This is evidenced in that up to 84\% of the 376,630 signals detected by the pipeline are removed in the first step (Table \ref{tab:cosmic_processing}) and only 0.5\% of the detected signals from real-time processing remained after applying the drift rate limit of $\pm$11.4 \, Hz \, s$^{-1}$.

After all steps of the pipeline, 160 signals over the three observations remained for visual inspection. None of the signals are consistent in time and were present in all antennas. We therefore conclude that there are no technosignatures toward the LTT~3780 system in this C-band data set. We have placed an upper limit on the \ac{EIRP} of 5.8 $\times 10^{12}$\,W (Table \ref{tab:eirp}).

\subsubsection{X-band Data}
\label{sssec:vla_x}

With the X-band receivers, we observed in two frequency segments, 8--8.8\,GHz and 8.9--10\,GHz. After applying a drift rate filter of $\pm$13.8\,Hz\,s$^{-1}$, removing known sources of \ac{RFI}, and applying a \ac{SNR} cut, 87\% of the signals detected by automated processing were removed. No signals were matched between the three days of observation in both drift rate and frequency, leaving the remaining 13\% as potential candidates. There were 18 signals that matched between coherent and incoherent beams and all had a unity ratio, suggesting that the signals were a form of local \ac{RFI}.

We therefore had 208 signals to be visually inspected. After evaluating the beamformed and individual antenna voltage data, no signals showed signals present in all antennas and persistent in time. At the end of the post-processing pipeline, no signals remained consistent with our definition of a technosignature. The upper limit of \ac{EIRP} is then 4.7 $\times 10^{12}$\,W, as listed in Table \ref{tab:eirp}.

\begin{deluxetable}{lccc}
\tablecaption{EIRP Limits for LTT 3780 c for each receiver band observation. \label{tab:eirp}}
\tablehead{
\colhead{Telescope} &
\colhead{Band} &
\colhead{Frequency Range (MHz)} &
\colhead{EIRP Limit (10$^{12}$ W)}
}
\startdata
VLA & S & 2307--3691 & 6.9 \\
VLA & C & 4807--6191 &  5.8\\
VLA & X & 8020--9998 &  4.7\\
ATA & L & 1000--1672  &  28\\
ATA & L, S & 1672--2344  &  30\\
ATA & S & 2344--3016  &  35\\
ATA & S & 3016--3688  &  36\\
\enddata
\tablecomments{EIRP limits correspond to the minimum detectable isotropic transmitter power at the distance of LTT~3780~c, using the sensitivity of each observing band and assuming narrowband emission. These values do not include a correction for de-smearing (i.e., $\beta$ from \citet{Gajjar_2021}.}
\label{tab:eirp_limits}
\end{deluxetable}

\subsection{Summary of results}
The 6,842 ATA hits and 546 VLA hits were subjected to visual inspection. Candidate signals were evaluated for temporal persistence, Doppler drift, spatial localization, and consistency across all contributing antennas. The strongest candidates were rejected as either single-antenna artifacts, persistent terrestrial interference, or signals simultaneously present in multiple synthesized beams. Consequently, no candidate exhibited the combined temporal, spectral, and spatial characteristics expected of a genuine astrophysical technosignature.

\section{Discussion}
\label{sec:discussion}

We conducted a search for technosignatures down to an EIRP limit sufficient to detect transmitters comparable to the Arecibo planetary radar ($\sim$20~TW; \citealt{Siemion_2013}). However, sensitivity expressed solely in terms of EIRP does not fully capture detectability. The efficacy of a technosignature search depends not only on the transmitter power but also on the temporal and spectral characteristics of the signal \citep{Gajjar_2026}. Our search strategy is optimized for persistent, narrowband, Doppler-drifting emission; however, extraterrestrial transmitters may instead exhibit intermittent, broadband, pulsed, or frequency-agile behavior. In such cases, detectability becomes a function of duty cycle, modulation scheme, and spectral occupancy rather than raw power alone.

The absence of detected signals in this study therefore constrains only a subset of possible technosignature parameter space—namely, continuous narrowband emitters within the searched drift-rate and frequency ranges. This non-detection does not preclude the presence of technological activity in the LTT~3780 system, but instead places limits on specific classes of transmitters under assumptions analogous to terrestrial radio technologies. In particular, Earth’s own radio leakage is time-variable and orders of magnitude weaker than high-power planetary radar systems \citep{saide2023simulation}, highlighting that even technologically active civilizations may remain undetectable under conventional narrowband search strategies.

A key strength of this work lies in the complimentary, multi-epoch observing strategy spanning multiple orbital phases, including transits and planet--planet occultations. 
Observations timed to these configurations are motivated by the possibility of enhanced detectability due to geometric alignment or an increased likelihood of intercepting intra-system communication signals. In particular, \citet{Fan_2025} estimate that planet--planet occultation geometries could enhance the probability of intercepting radio spillover from intra-system communication by factors as large as $\sim4\times10^5$ in the analogous Earth--Mars case.

The complementary ATA and VLA observations demonstrate the value of applying multiple observational and processing strategies to the same planetary system. As discussed in 
Section \ref{sec:intro}, the two facilities probe overlapping technosignature parameter space through distinct beamforming and candidate-identification methodologies, reducing dependence on any single pipeline architecture.


More broadly, this study demonstrates the feasibility and scientific value of integrating technosignature searches into coordinated, multi-wavelength investigations of exoplanetary systems. In the case of LTT~3780, recent JWST observations indicating a chemically active atmosphere provide strong motivation for such joint studies. Although atmospheric detections alone cannot uniquely distinguish between biological, geological, or technological processes, technosignature searches offer an independent and complementary probe of planetary activity. The absence of detectable radio technosignatures in this system therefore provides an additional constraint that can be interpreted alongside atmospheric composition and planetary context.

Future work will benefit from expanding both temporal coverage and search methodologies. Repeated observations across a wider range of orbital phases and longer baselines will improve sensitivity to intermittent or geometry-dependent signals. In parallel, the development of search algorithms targeting broader classes of emission—such as transient, broadband, or modulated signals—will be essential to investigate the parameter space of the technological signature beyond the narrowband paradigm. Together, these efforts will enable for more comprehensive constraints on technological activity in nearby planetary systems.

\acresetall

\section{Conclusion}
\label{sec:conclusion}
We conducted complementary technosignature searches toward the exoplanetary system LTT~3780 using the ATA and the VLA, spanning multiple orbital phases and covering 1–10\,GHz. No candidate signals consistent with narrowband, Doppler-drifting technosignatures were identified after comprehensive RFI mitigation and multi-beam verification. We place upper limits on the minimum detectable isotropic transmitter power of $4.7\times10^{12}$–$3.6\times10^{13}$\,W across the observed bands, corresponding to sensitivities capable of detecting transmitters comparable to or stronger than the Arecibo planetary radar at the distance of LTT~3780. Although no technosignatures were detected, this work demonstrates the value of coordinated, multi-facility observations aligned with known orbital configurations and illustrates a scalable framework for integrating technosignature searches into broader exoplanet biosignature studies. Continued observations across diverse time–frequency parameter space and expanded search methodologies will further refine constraints on technological activity in nearby planetary systems.

\section{Acknowledgments}
\label{sec:acknowledgments}

The authors acknowledge the foundational support from John and Carol Giannandrea that made COSMIC possible. The National Radio Astronomy Observatory is a facility of the National Science Foundation operated under a cooperative agreement with Associated Universities, Inc. The Allen Telescope Array (ATA) refurbishment program and its ongoing operations have received substantial support from Franklin Antonio. Additional contributions from Frank Levinson, Greg Papadopoulos, the Breakthrough Listen Initiative, and other private donors have been instrumental in the renewal of the ATA. The Paul G. Allen Family Foundation provided major support for the design and construction of the ATA, alongside contributions from Nathan Myhrvold, Xilinx Corporation, Sun Microsystems, and other private donors. The ATA has also been supported by contributions from the US Naval Observatory and the US National Science Foundation. Breakthrough Listen is managed by the Breakthrough Initiatives, sponsored by the Breakthrough Prize Foundation.

\bibliography{cosmic}{}
\bibliographystyle{aasjournalv7}

\section{Data Availability}
\label{sec:data_availability}

All data were collected via the commensal backend on the VLA telescope (COSMIC) and the Allen Telescope array. Both are under the ownership of Breakthrough Listen and the SETI Institute. The voltage, filterbank, or database information around each hit can be made available upon request to the authors. It is our plan in the future to make this data publicly available through a web service.

\begin{acronym}
    \acro{SETI}{Search for Extraterrestrial Intelligence}
    \acro{ETI}{Extraterrestrial Intelligence}
    \acro{ATA}{Allen Telescope Array}
    \acro{RFI}{Radio Frequency Interference}
    \acro{SNR}{Signal-to-Noise Ratio}
    \acro{FRB}{Fast Radio Burst}
    \acro{BLADE}{Breakthrough Listen Accelerated Digital signal processing  Engine}
    \acro{FWHM}{full width at half maximum}
    \acro{BL}{the Breakthrough Listen Initiative}
    \acro{HCRO}{Hat Creek Radio Observatory}
    \acro{FCC}{Federal Communications Commission}
    \acro{SEFD}{System Equivalent Flux Density}
    \acro{EIRP}{Effective Isotropic Radiated Power}
    \acro{VLA}{Very Large Array}
    \acro{PPO}{Planet-Planet Occultation}
    \acro{ATSAT}{the Allen Telescope Survey Agnostic Technosignature Pipeline}
    \acro{NRAO}{National Radio Astronomy Observatory}
\end{acronym}

\appendix
For each observatory, there are a series of channels that contain known radio frequency interference. The tables displayed in this Appendix outline which channels were flagged prior to analysis. This is because any signal detected would be suspect as it coincides with known Earth-related frequency emitters. Tables \ref{tab:blanking_ranges_cosmic} and \ref{tab:blanking_ranges} outline which channels each telescope is considered to contain a strong RFI.

\begin{deluxetable*}{lllllll}
\tablecaption{The frequency-blanking ranges, in MHz, used for the \ac{ATA} survey.\label{tab:blanking_ranges}}
\tablewidth{0pt}
\tablehead{}
\startdata
1000--1025   & 1525--1562 & 1951--1980     & 2067.3--2067.8 & 2237--2241   & 2353--2359.5    & 3341--3343 \\
1075--1116   & 1572--1580 & 1990--1996     & 2097--2098     & 2244--2248   & 2484--2488      & 3492--3493\\
1168--1184.3 & 1671--1705 & 1999--2000.5   & 2110--2154.6   & 2257--2258   & 2497.25--2498.5 & 3600--3688\\
1187--1227   & 1775--1802 & 2010--2011     & 2166--2169.5   & 2269--2270.3 & 2565.75--2585   &\\
1246--1250   & 1811-1813 & 2022-2023     & 2179--2203.5   & 2271.3--2279 & 2590--2609      &  \\
1275--1314   & 1899--1901 & 2049--2054     & 2211--2212     & 2299--2301   & 2705--2800      & \\
1395--1420.5 & 1927--1945 & 2062.3--2062.8 & 2226--2227.5   & 2316--2345   & 3264--3274      & \\
\enddata
\tablecomments{These ranges comprise a cumulative bandwidth of 734.4~MHz. Any \texttt{bliss} output with center frequency within a blanking range was removed due to high levels of \ac{RFI} in these regions.}
\end{deluxetable*}

\begin{deluxetable*}{lllllll}
\tablecaption{Frequency-blanking ranges (MHz) used in this work for the COSMIC Survey. \label{tab:blanking_ranges_cosmic}}
\tablewidth{0pt}
\tablehead{}
\startdata
2106--2107 & 2178--2195 & 2180--2290 & 2204--2205 & 2227--2231.5 & 2246--2252.5 & 2268.5--2274.5 \\
2282.5--2288.5 & 2314.5--2320.5 & 2320--2332.5 & 2324.5--2330.5 & 2332.5--2345 & 2334.5--2340.5 & 2387.5--2388 \\
2400--2483.5 & 2411--2413 & 2483--2500 & 2741--2742 & 2791--2792 & 3700--4200 & 5648.5--5663.5 \\
5659.5--5670.5 & 5695.5--5704.5 & 5742.5--5757.5 & 5765--5769 & 5770--5880 & 5796--5804 & 6020--6055 \\
6108.1--6138.1 & 6110--6120 & 6121--6240 & 6137.75--6167.75 & 6182--6212 & 6270--6305 & 6360.14--6390.14 \\
6360--6420 & 6389.79--6419.79 & 6772--6778 & 7250--7850 & 9300--9300 & 10740--10740 & 10820--10820 \\
10957--10957 & 11037--11037 & 11230--11230 & 11310--11310 & 11447--11447 & 11527--11527 & 11700--11700 \\
\enddata
\end{deluxetable*}

\end{document}